\documentclass[11pt]{article}

\input epsf.sty

\def\lromn#1{\uppercase\expandafter{\romannumeral#1}}

\usepackage{graphicx,amsmath,amssymb}
\def\lromn#1{\uppercase\expandafter{\romannumeral#1}}

\begin{document}
\begin{flushright}
\today \\
%${\cal M. Y.}$
\end{flushright}

\vspace{2cm}
\begin{center}
\begin{Large}
{\bf Two-photon paired solitons supported by
medium polarization
}

\end{Large}

\vspace{1cm}
%\begin{center}
\begin{large}
M. Yoshimura and N. Sasao$^{\dagger}$

Center of Quantum Universe, Faculty of
Science, Okayama University \\
Tsushima-naka 3-1-1 Kita-ku Okayama
700-8530 Japan\\

$^{\dagger}$
Research Core for Extreme Quantum World,
Okayama University \\
Tsushima-naka 3-1-1 Kita-ku Okayama
700-8530 Japan 
\end{large}
\end{center}

\vspace{4cm}

\begin{center}
\begin{Large}
{\bf ABSTRACT}
\end{Large}
\end{center}

We derive for the first time fundamental equations
that describe soliton spatial profiles
consisting of two-photon mode fields and
macroscopic polarization of medium.
Numerical solutions of this basic equation
are presented to suggest both single soliton and multiple
soliton chains in  infinitely long targets, 
taking an example of para-H$_2$ $v=0 \leftrightarrow 1$ 
(E1 forbidden) vibrational parameters. 
Although effects of dissipative relaxation are included
in the general form for the two-level system,
the existence of static soliton-condensate is established.
For finite size targets we can precisely formulate
the profile equation in  a framework of non-linear eigenvalue problem.
Its first iteration provides approximate semi-analytic results
under a potential well in the linearized equation,
which have qualitatively similar profiles to the case of
infinitely long target, an important difference
being the exponentially decreasing profile near target ends proper to
the eigenvalue problem.
A large number of weakly interacting solitons correspond
to localized portions between adjacent nodes of highly excited bound  state
wave functions  in a one-dimensional potential well of large size.
These soliton-condensates are expected to
be important to enhance the signal to the background ratio in the proposed neutrino
mass spectroscopy using atoms.

\vspace{2cm}

Key words

Paired super-radiance,
Polariton,
Soliton,
Non-linear Schr\"odinger equation,
Neutrino mass spectroscopy

\newpage

\newpage
\lromn1 
{\bf Introduction}

Solitons of macroscopic size which are made of  dynamical
electromagnetic fields supported by
medium polarization  are of great interest
both from points of fundamental physics and applications.
Basic entity when medium atoms are
excited by the usual electric dipole transition
is known as polariton \cite{hpfield}, \cite {polariton overview}.
The study of solitons made of the 
dipole field has a long history
since the early 
works \cite{coherent light propagation in 2 level}.
Polariton type of solitons are expected to
play fundamental roles in a new type
of lasing material and quantum information,
and there have been remarkable experimental
progress on polariton condensates 
\cite {polariton exp 1}, \cite {polariton exp 2}.

We investigate in the present work
solitons of a different kind;
the system of medium atoms that are coupled to
two-photon mode.  The transiton between two relevant atomic
levels are E1 forbidden like
$J=2 \rightarrow 0, \Delta P = +$ transitions,
but E2 or two-photon allowed.
Time evolving macro-coherent dynamical process
$|e \rangle \leftrightarrow |g\rangle + \gamma \gamma$ has been studied
in our previous works \cite{yst pra}, \cite{ptep overview}
and existence of explosive two-photon events
of characteristic features have been established.
This dynamical process is termed paired super-radiance (PSR).
 Highly entangled two-photon states that
emerge in PSR  may have an advantage  over the usual type of polariton
in some aspects of applications.
Moreover, at the fundamental level
PSR dynamics is closely related to 
the proposed macro-coherent radiative emission
of neutrino pair (RENP) 
$|e \rangle \rightarrow |g\rangle + \gamma+\nu \nu$
\cite {ptep overview}, \cite{ys-13}
 when an applied static electric field  
induces admixture of different parity state appropriate
for RENP.
Once soliton-condensates related to PSR are formed,
they may become an ideal target state of RENP, because
these condensates are  stable remnants and 
do not emit QED background two-photon pairs from target ends.
Many advantages of atomic RENP for future
neutrino physics have been reviewed in \cite{ptep overview}.

In the present work we derive for the first time the
correct form of soliton profile equations related to PSR.
Effects of phase-coherence relaxation and population decay
are both included in this derivation.
By numerical calculations we show that both
single soliton and multiple soliton chains are (almost,
in an approximate sense)
derived by solutions of this profile equation in an infinitely long medium.
Results are surprising in the sense that
these stable soliton solutions exist despite of
the most general form of dissipative terms
in the two-level system.
For finite-size targets we are able to formulate
a non-linear eigenvalue problem and
bound states of the eigenvalue problem 
correspond to soliton-condensates.
The existence  of stable soliton-condensates
is guaranteed by a form of one-dimensional potential well
that appears in the first iteration of
linearized approximation.
To the best of our  knowledge, it is rare or none that
soliton solutions in the presence of dissipation
are explicitly shown.

Significance of soliton formation in the neutrino
mass spectroscopy is in enhancing the signal to the background 
raito and in providing an ideal  experimental means.
For this experimental purpose
 a fast and stable switching mechanism between PSR and
RENP modes has to be devised.

The rest of the present work is organized as follows.
In Section \lromn2, after a brief review of
the modified master equation of PSR dynamics
we derive  the static profile equation for envelopes of
field components after the Bloch vector is solved
in terms of field components.
With the symmetry of PSR system fully taken into
account we reduce the effective field degrees
of freedom to two real fields, which
may be taken as two counter-propagating field
strengths.
A classical mechanical analogue of
particle in two space dimensions is pointed out.
The acting force is not conservative.
In Section \lromn3 the soliton profile equation
thus derived is numerically solved  for infinitely long targets.
Both single and multiple soliton
chain solutions are identified apprximately.
We present numerically obtained profiles of soliton-condensates
for para-H$_2$ vibrational transition $v=0 \leftrightarrow 1$,
a system strictly E1 forbidden with the main de-excitation mode
being two-photon process.
In Section \lromn4 the finite size of targets
is treated and a precise non-linear eigenvalue problem
is formulated.
The first order iteration of this non-linear eigenvalue
problem is reduced to a linearized potential well
problem in one space dimension.
Its wave functions of bound states give condensate profiles,
the probability amplitudes corresponding to
field magnitudes precisely.
For a potential well given by a large coherent region
of one-to-one (1/1) mixture of excited and ground states, highly excited $n-$th level is
well described by the WKB formula of quantum mechanics
with $n-1$ nodes.
A $n$ number of weakly interacting solitons in the condensate 
may thus be identified by  localized objects between two adjacent nodes.
In Appendix we give some technical details
on the master equation and input data of para-H$_2$.

Throughout this work
we use the natural unit of $\hbar = c = 1$.

\vspace{0.5cm}
\lromn2 
{\bf Derivation of soliton profile equation}

Since we assume that the target is irradiated 
by excitation and trigger
lasers in one and its counter-propagating
direction, the development of medium polarization
occurs only in a direction taken here as $x$ axis.
We further assume that target atoms
are distributed uniformly and take
the continuous limit of target atom distribution.
A further simplification we adopt for
the moment is to take
an infinitely large target.

The main part of two-photon interaction with
two level atoms is described by an effective hamiltonian
in 1+1 dimensional field theory:
\begin{eqnarray}
&&
H = \int dx E^2(x)
\left( c_e^*(x) , c_g^*(x)
\right)
\alpha
\left(
\begin{array}{c}
c_e(x)  \\
c_g(x)   
\end{array}
\right)
\,,
\end{eqnarray}
where $c_i(x)$ are wave functions for  atoms
in the level $|i \rangle$
and $\alpha$ are $2\times 2$ matrix given by product of two
dipole transition elements $d_{pe}, d_{pg}$ 
to an upper intermediate level $|p\rangle$ 
coupled to two lower levels $|e \rangle, |g\rangle$ by E1 dipole
moments \cite{yst pra}.
The hamiltonian is given in an interaction picture,
hence the off-diagonal elements contain oscillating
functions $e^{\pm i \epsilon_{eg} t}$
which may be eliminated by slowly varying
envelope approximation (SVEA) done later.
This hamiltonian is derived by taking the Markov approximation
and eliminating $|p\rangle $.  
Numerical parameters of $\alpha$ matrix elements are
illustrated for the example of para-H$_2$ vibrational system of
$v=1 \leftrightarrow 0$ in Appendix.
In a typical situation the field $E(x) = E_R(x) + E_L(x)$
is decomposed into a sum of right and left moving waves
of different frequencies whose sum coincides with the
energy difference, $\omega_1 + \omega_2 = \epsilon_{eg}$.

From this hamiltonian one may derive the Bloch equation for
bilinear forms of
wave functions, namely
the density matrix elements, 
$R_i = \psi^{\dagger} \sigma_i \psi$
(with the normalization of the number density $n$ for  $\psi^{\dagger}\psi$).
We include the most general dissipation term
given by a rigorous analysis \cite {lindblad} for the two-level system,
which agrees with description in terms of $T_1, T_2$ due to Bloch.
The medium polarization $P$ is calculated from this density matrix element
and inserted in the right hand side (RHS) of Maxwell equation,
$(\partial_t^2 - \partial_x^2) E = - \partial_t^2 P$.
The resulting system of non-linear partial differential equations 
for dynamical
variables, $R_i, E_i$, is called Maxwell-Bloch equation and constitutes
the master equation for PSR dynamics after matched phase components are 
projected out.
The master equation after reduction to
the first order differential equations with
slowly varying envelope approximation (SVEA) 
is given in \cite{ptep overview}.

We seek static solutions of the Maxwell-Bloch equation.
An important departure taken here from \cite{ptep overview} is that 
we rely on second order differential equations of
space-time derivatives.
This is because experience in known soliton solutions
such as the real field Goldstone model \cite{coleman}
shows that the second order formalism is essential
for derivation of solitons.
With the spirit of SVEA we replace the derivative operation
in RHS of Maxwell equation by $-\partial_t^2 \rightarrow
\omega_i^2$.
The resulting new master equation for PSR is given
in Appendix.
We seek static remnant solutions of this master
equation.

An important change in the second order formalism
is in basic scale units of time, length and field strength.
They are given by
\begin{eqnarray}
&&
t_0 = (\frac{1}{2} \epsilon_{eg} \sqrt{\alpha_{ge}n})^{-1}
\,, \hspace{0.5cm}
l_0 = c t_0
\,, \hspace{0.5cm}
E_0^2 = \epsilon_{eg} \sqrt{\frac{n}{\alpha_{ge}}}
\,.
\end{eqnarray}
We refer to Appendix for precise definition of $\alpha_{ge}$.
These differ from $t_*, E_*^2$, the units defined in 
\cite{ptep overview} (reflecting the  change from the first to the second
order formalism of field equations), in the important 
$n$ (target number density) dependence.
For the para-H$_2$ transition, $ct_0 \sim 0.03$mm with $n=10^{21}$cm$^{-3}$
scaling like $1/\sqrt{n}$.
We define dimensionless length ($\xi$) and time ($\tau$) as
$\xi = x/l_0\,, \tau = t/t_0$.

The static version of Bloch equation is solved in the presence of inhomogeneous term 
(which reflects that the system ultimately approaches  the
ground state), 
resulting in the Bloch vector components
written in terms of fields.
This is given by an $3\times 3$ matrix inversion for $\vec{r} = \vec{R}/n$,
\begin{eqnarray}
&&
\vec{r} = {\cal R}^{-1} 
\left(
\begin{array}{c}
0  \\
0 \\
1
\end{array}
\right)
\,, \hspace{0.5cm}
{\cal R} = \tau_1 {\cal A}
- \left(
\begin{array}{ccc}
\tau_{1/2} & 0 & 0\\
0 & \tau_{1/2} & 0\\
0 & 0 & 1
\end{array}
\right)
\,, \hspace{0.5cm}
\tau_{1/2} = \frac{T_1}{T_2}
\,, \hspace{0.5cm}
\tau_{1} = \frac{T_1}{t_0}
\,,
\end{eqnarray}
where ${\cal A}$ is an anti-symmetric $3\times 3$ matrix,
\begin{eqnarray}
&&
{\cal A} =
\left(
\begin{array}{ccc}
0 & a & -b \\
-a & 0 & c\\
b & -c &  0
\end{array}
\right)
\,,
\hspace{0.5cm}
a =2 \gamma_-(|e_R|^2 + |e_L|^2) 
\,, \hspace{0.5cm}
b= - 4  \Im (e_R e_L)
\,, \hspace{0.5cm}
c= 4 \Re (e_R e_L)
\,,
\end{eqnarray}
where $e_i, i=R,L$ are dimensionless fields 
in the unit of $E_0$
of right and left moving field envelopes (with the factor $e^{\pm i \omega t}$ taken out)
and $\gamma_{\pm} = (\alpha_{ee} \pm \alpha_{gg})/2\alpha_{ge}$.
In the para-H$_2$ example, $\gamma_- \sim 0.64\,, \epsilon_{eg} \sim
0.52 $eV, and $\epsilon_{eg} n \sim $ 26 GW mm$^{-2}$ for 
$n= 10^{21}$cm$^{-3}$.
The resulting Bloch vector is inserted into
RHS of Maxwell equation and
the soliton profile equation is formulated,
using real fields.
The  profile equation for real variables is given by
\begin{eqnarray}
&&
\vec{e}^T =
\left(
\Re e_R\,, \Im e_R  \,,
\Re e_L  \,,
\Im e_L
\right) \equiv (e_1, e_2, e_3, e_4)
\,,
\\ &&
- \frac{d^2}{d\xi^2}\vec{e}
= {\cal M} \vec{e}
\,,\hspace{0.5cm}
{\cal M} =
\frac{1}{2}\left(
\begin{array}{cc}
 \gamma_+ + \gamma_-r_3  & r_1 \sigma_3 - r_2 \sigma_1 \\
r_1 \sigma_3 - r_2 \sigma_1 & \gamma_+ + \gamma_- r_3 
\end{array}
\right)
\,,
\label {profile eq in e}
\\ &&
r_1 = - \frac{4\tau_2}{D}
\left( \Im (e_R e_L) + 2 \tau_2 \gamma_-
\Re (e_R e_L)(|e_R|^2 + |e_L|^2)
\right)
\,,
\\ &&
r_2 = - \frac{4\tau_2}{D}
\left( \Re (e_R e_L) - 2 \tau_2 \gamma_-
\Im (e_R e_L)(|e_R|^2 + |e_L|^2)
\right)
\,,
\\ &&
r_3 = -\frac{1 + 4 \gamma_-^2 \tau_2^2
(|e_R|^2 + |e_L|^2)^2}{D}
\,, \hspace{0.5cm}
D = 1 + 4 \gamma_-^2 \tau_2^2
(|e_R|^2 + |e_L|^2)^2 
+ 16 \tau_1 \tau_2 |e_R e_L|^2
\,,
\end{eqnarray}
in the $2\times 2$ block-diagonal form.
$4\times 4$ matrix ${\cal M}$ is real and symmetric.
The middle point of frequency $\omega_i = \epsilon_{eg}/2\,, i=1,2$ is taken
for simplicity.
We do not expect qualitatively different results for
another choice of $\omega_i$.

It is useful to think of classical mechanical
analogy by taking $\vec{e}$ as 4-vector of particle position,
$\xi$ as a fictitious time
and RHS of (\ref{profile eq in e})
as a force.
This force is not conservative, since
$\partial_i ({\cal M}_{jk} e_k) \neq \partial_j ({\cal M}_{ik} e_k)$.

Eigenvalues of the matrix ${\cal M}$  play important roles and
it turns out that these are functions of 
$R = |e_R|^2$ and $L = |e_L|^2$ alone.
We may thus expect that classical mechanics of 
this system is governed by these two variables
and that there exist two-fold symmetry.
This symmetry is associated with two conserved quantities,
\begin{eqnarray}
&&
W' = e'_1 e_2 - e'_2 e_1 -e'_3 e_4 + e'_4 e_3 
\,, \hspace{0.5cm}
XY' - YX'
\,,
\end{eqnarray}
with $X = \Re(e_R e_L), Y = \Im (e_R e_L)$.
$'$ indicates $\xi$ derivative.

Imposing these conservation makes it possible to write
the soliton profile equation as
ordinary second order differential equations
in terms of right and left moving fluxes $R, L$:
\begin{eqnarray}
&&
R'' = \frac{(R')^2}{2R} - 
\left( \gamma_+ +  \gamma_- r_3
\right)R  + \frac{2 (l-hR)^2}{R}
\,,
\label {soliton profile eq r}
\\ &&
L'' = \frac{(L')^2}{2L} - 
 \left( \gamma_+ +  \gamma_- r_3
\right) L   + \frac{2 (l+hL)^2}{L}
\,,
\label {soliton profile eq l}
\\&&
r_3 = - \frac{ 1+ 4\tau_2^2\gamma_-  ( R +  L )^2  }
{ 1 + 16 \tau_1 \tau_2 RL+ 4\tau_2^2 \gamma_-^2 
( R +  L )^2 }
\,,
\end{eqnarray}
with two conserved quantities $h,l$ given by
\begin{eqnarray}
&&
XY' - YX' = l (R+L)
\,, \hspace{0.5cm}
W' = h (R+L)
\,.
\end{eqnarray}
Physical meaning of these variables, 
 $l$ and and $h$, are angular momentum
components for the particle motion in four dimensions
in appropriate units when one regards space coordinate
$\xi$ as a time.

We have neglected a small term 
\begin{eqnarray}
&&
\frac{8\tau_2^2 \gamma_-}{ 1 + 16 \tau_1 \tau_2 RL+ 4\tau_2^2 \gamma_-^2 
( R +  L )^2 }RL(R+L)
\,,
\end{eqnarray}
in RHS of these equations that vanishes in
the $T_1 \gg T_2$ limit, which is valid in most targets.
In numerical computations in the next section
this small term is however included.

\vspace{0.5cm}
\lromn3 
{\bf Condensate profiles and its interpretation}

We first give a variety of numerical solutions of
the full soliton profile equation, 
eqs(\ref{soliton profile eq r}), (\ref{soliton profile eq l}).
Since our final equations are a coupled set of non-linear ordinary
differential equations for two components,
numerical results are reliable, and we
are able to identify important classes of solutions
sufficient to clarify the overall picture of static solutions.
We impose the boundary condition at the center of target
such that field derivatives vanish there with a variety
of field value choices.
Since we do not impose the appropriate condition for the vacuum
outside targets, we are effectively treating an infinitely long
target in most of discussions that follow.
Next we give interpretation of these general solutions
in terms of aggregate of single solitons that have
topological property \cite {coleman}.

The useful quantity that gives directly the 
potential activity factor of both PSR and RENP is
\begin{eqnarray}
&&
\frac{d\eta}{d\xi} =
(r_1^2 + r_2^2) (|e_R|^2 + |e_L|^2) =\frac{ 8\tau_2^2RL(R+L)}
{\left(  1 + 16 \tau_1 \tau_2 RL+ 4\tau_2^2 \gamma_-^2 
( R +  L )^2\right)^2}
\left( 1+ 4\tau_2^2  \gamma_-^2 (R +L)^2
\right) 
\,.
\label  {eta integrand}
\end{eqnarray}
This quantity is  shown in some presented figures below.
The largest integrated value $\eta = \int d\xi d\eta/d\xi$ of this quantity is 1.
Corresponding to this largest value 
RENP rate (proportional to the activity factor $\eta$) becomes maximal.
For instance, in the  example of Cs $P_{1/2}$(1.39 eV) de-excitation studied 
in relation to the nuclear monopole pair emission
\cite{ys-13}
the rate becomes of order
$10^4$Hz for the target number density $n=10^{21}$cm$^{-3}$
(rate proportional to $n^3$)
and for the target volume $V= 10^2$cm$^3$
(proportional to $V$).

\begin{figure*}[htbp]
 \begin{center}
 \epsfxsize=0.6\textwidth
 \centerline{\epsfbox{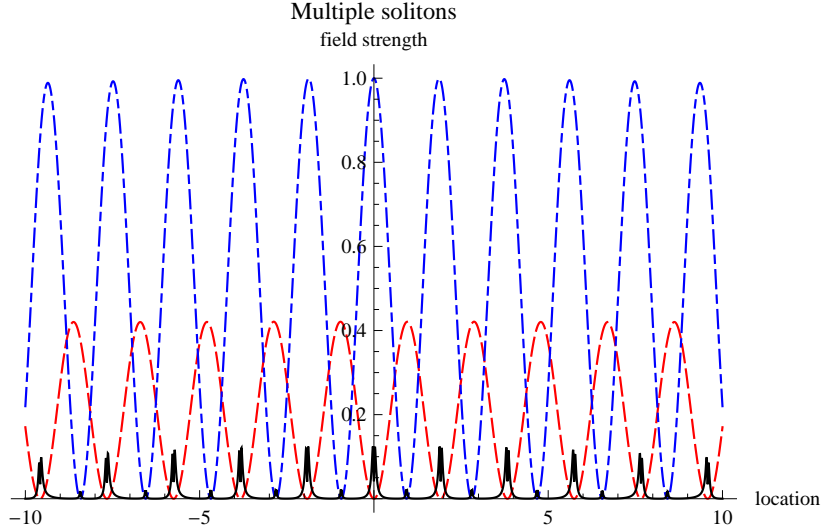}} \hspace*{\fill}
   \caption{Para-H$_2$ PSR soliton solution for
$(h, l) = (-1,0.01)$ under the condition of
$R=10^{-4}, L=1, R'=L' =0$ at the center of target. R-mover energy density in 
dashed red, L-mover in dash-dotted blue and
$\eta-$profile times 50 in solid black.
Assumed parameters are $\tau_1 = 1000, \tau_2=10$.
}
   \label {soliton profile 4}
 \end{center} 
\end{figure*}

\begin{figure*}[htbp]
 \begin{center}
 \epsfxsize=0.6\textwidth
 \centerline{\epsfbox{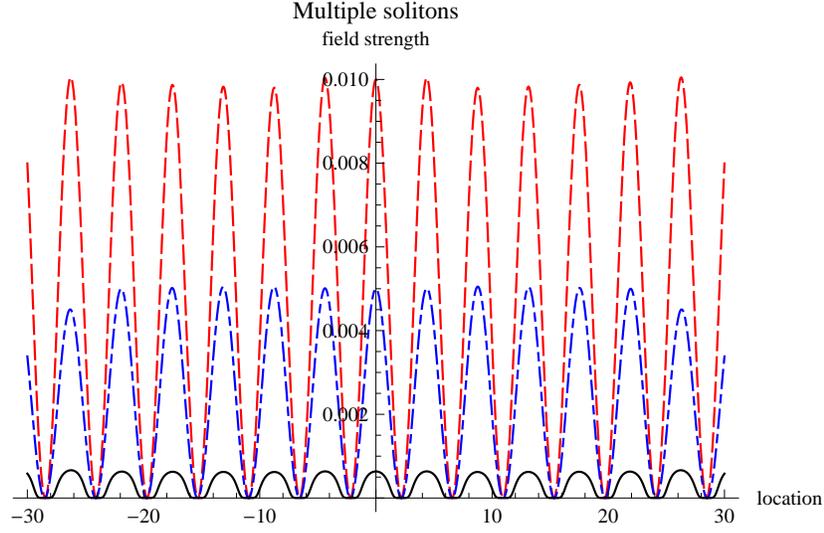}} \hspace*{\fill}
   \caption{Example of strongly interacting soliton-condensates.
$(h, l) = (-1.8, 0)$ with the boundary condition of
$R=0.01, L=0.005, R'=L' =0$ at the center of target. R-mover energy density in dashed red, 
L-mover in dash-dotted blue and
$\eta-$profile times 100 in solid black.
The unit length 1 in this figure corresponds to $\sim$ 0.1 mm
for para-H$_2$ atom density $10^{21}$cm$^{-3}$. 
Assumed relaxation parameters are $
\tau_1 = 1000, \tau_2=10$.
}
   \label {ph2 soliton profile}
 \end{center} 
\end{figure*}

\begin{figure*}[htbp]
 \begin{center}
 \epsfxsize=0.6\textwidth
 \centerline{\epsfbox{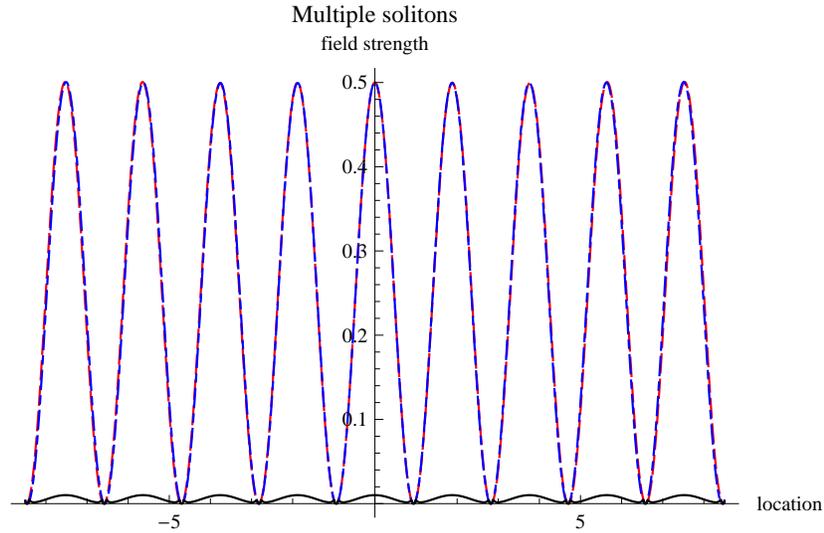}} \hspace*{\fill}
 \caption{Soliton-condensate for a finite size target.
$R-$mover energy density in dashed red,
$L-$mover in dash-dotted blue which is degenerate with $R-$mover.
$500 \times d\eta/d\xi$ is plotted in solid black
Assumed parameters are $h= -1, l=0.001,
\tau_1 = 10^3, \tau_2=10$.
}
   \label {soliton of finite size} 
 \end{center} 
\end{figure*}

\begin{figure*}[htbp]
 \begin{center}
 \epsfxsize=0.6\textwidth
 \centerline{\epsfbox{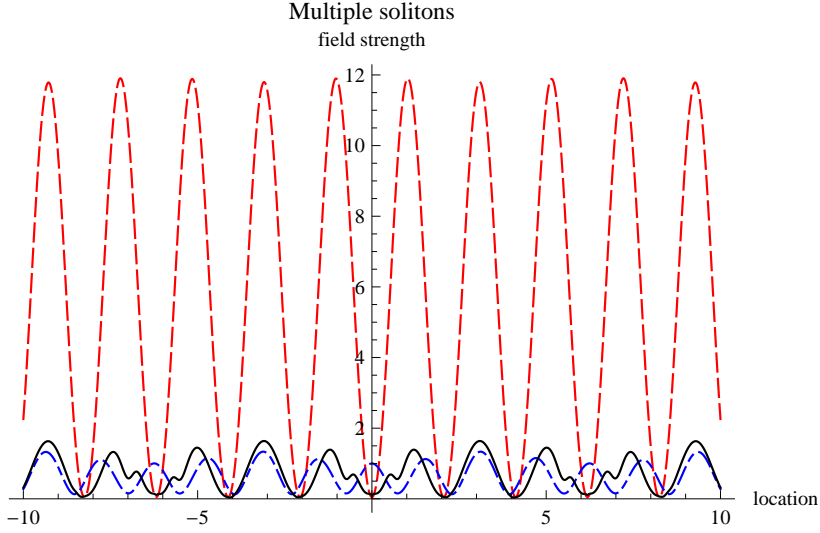}} \hspace*{\fill}
 \caption{Example of large activity factor $\eta$.
R- and L-mover energy density in dashed red and in dash-dotted
blue respectively. The profile $d\eta/d\xi$ is depicted in solid black
(without any multiplicative factor).
Assumed parameters are $(h,l) = (-1,1) \,, \tau_1=\tau_2 = 10$ and 
the boundary condition is taken as
$R=0.05 , L=1, R'=L' = 0$ at the center.
}
   \label {large eta case} 
 \end{center} 
\end{figure*}

A typical feature of many solutions such as shown
in Fig(\ref {soliton profile 4}) ,  Fig(\ref {ph2 soliton profile}),  
Fig(\ref {soliton of finite size}) and Fig(\ref {large eta case}),  is
a periodic structure which may be decomposed into
two types of basic units, one unit made  of large L-mover energy density and
small R-mover energy density at one end and a
large R-mover and small L-mover energy densities at another end.
The other unit is made of L/R ratio reversed from this one.
This class of solutions have been obtained for  large $h$ of order unity
and small $l$ except the case of Fig(\ref {large eta case})
in which both $h,l$ are of order unity.
The structure of these profiles suggests that
static solutions are  aggregates of these two basic units of
solitons.
In \cite{yst pra} these two type of  solitons are called
emitter and absorber solitons and
they are characterized by opposite topological
quantum numbers of $\pm 1$.
Emitter and absorber are arranged alternatively in Fig(\ref {soliton profile 4}), hence
may be designated by $\cdots E-A-E-A \cdots$ chain.

At two target ends of these one-dimensional objects
both R- and L-moving energy densities are non-vanishing in general,
which means that their energy fluxes are also non-vanishing.
Thus, these objects truncated
from the infinitely large target to a finite length target are expected to
be unstable due to the energy leakage from two ends.
Flux leakage can  be ignored for a large enough target, and one
has the picture of aggregate of stable soliton chains
albeit in a good approximate sense.

\begin{figure*}[htbp]
 \begin{center}
 \epsfxsize=0.6\textwidth
 \centerline{\epsfbox{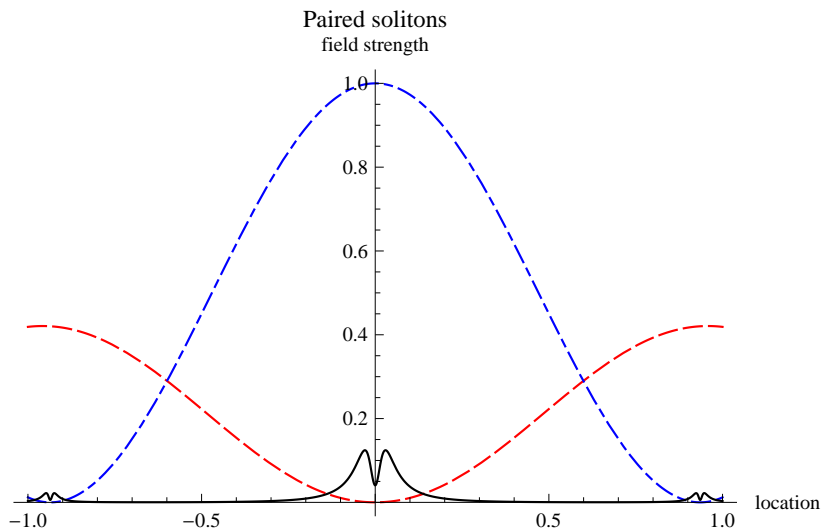}} \hspace*{\fill}
   \caption{Portion of a paired solitons taken from
Fig(\ref {soliton profile 4}).
The left portion is absorber soliton and the right is emitter soliton.
}
   \label {soliton profile 5}
 \end{center} 
\end{figure*}

A single unit of emitter and absorber solitons
may be defined by a finite portion of solutions in
the ideal infinite size target, whose end points
are defined by two adjacent locations of 
vanishing derivatives, $R' = 0, L' = 0$.
These are the left and the right portions of
objects given in Fig(\ref {soliton profile 5}).
The left portion is absorber soliton, and the right
is emitter soliton.
The other pair of emitter in the left and absorber in the right
may also be identified.
Solitons thus obtained have finite sizes, and the infinite number of
these pairs appear in
static solutions for the infinite size target.

\begin{figure*}[htbp]
 \begin{center}
 \epsfxsize=0.6\textwidth
 \centerline{\epsfbox{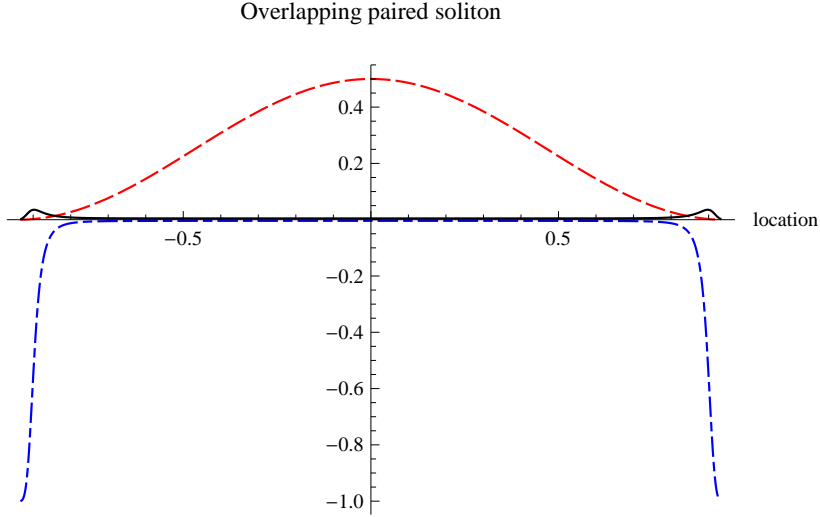}} \hspace*{\fill}
 \caption{Truncated portion corresponding to Fig(\ref{soliton of finite size})
that gives a single soliton of overlapping R and L energy
densities. Energy density is depicted in dashed red, 
$r_3$ in dash-dotted blue, and $\sqrt{r_1^2+r_2^2}$ in solid black.
}
   \label {single soliton} 
 \end{center} 
\end{figure*}

The result shown in figures of Fig(\ref{soliton profile 4}) and  Fig(\ref{soliton profile 5})
indicates that interaction of solitons is weak, hence the profile structure
of aggregate can readily be interpreted as a chain of ideal AE solitons.
But there are solutions that show much stronger interactions.
In Fig(\ref{ph2 soliton profile}) we show one of these solutions.
This example indicates that emitter-emitter or absorber-absorber
interaction is attractive and strong.
In this case field maxima of R- and L-movers appear at the same
target locations.
The extreme case of this class of solutions is given by
Fig(\ref{soliton of finite size}) in which R- and L-energy densities
coincide.
Its portion of a single peak is shown in Fig(\ref {single soliton})
where both eneregy density and Bloch vector components
are depicted.
The single peak solution is characterized by
a mixture of excited and ground state of
the ideal 1/1 mixing except the sharp edge region.

So far we showed results for infinitely long targets.
For finite size targets one imposes the condition that
R- and L-mover energy densities vanish at two ends
and assumes that they are matched to the trivial
vacuum solution outside the finite target.
In this case field maxima within the target cannot take
arbitrary given values, and these values become
quantized.
Thus, finite size soliton-condensates are obtained as solutions
to a kind of eigenvalue problem.
We shall formulate more precisely a non-linear eigenvalue problem in the next section,
but for a sufficiently long target our numerical solutions for
the infinitely long target give good approximate solutions to
the eigenvalue problem.
Indeed, the result given by Fig(\ref {soliton of finite size})
is a good approximate solution to the eigenvalue problem.
Its truncation given by Fig(\ref {single soliton}) is a bound pair of
solitons and of finite size.
Its size is of order a few times the basic length scale $ct_0$.

Finally, we show the case of large activity factor $\eta$ of order unity
in Fig(\ref {large eta case})  in which a different $\tau_2/\tau_1$ ratio
and $h,l $ values are taken.

It should be clear that our solutions for the infinitely long target
provide a variety of cases of soliton-soliton interaction.
Further clarification of soliton interaction is however
beyond the scope of the present work.

\vspace{0.5cm}
\lromn4 
{\bf 
Formulation of non-linear eigenvalue problem and
approximate semi-analytic results of soliton-condensates
}

We first give a potential well model based on
heuristic arguments along with its semi-analytic results, and later formulate
the non-linear eigenvalue problem.
Solutions in the first part are found to be
the first order approximate solutions to the precise non-linear problem
in the second part; the first one in a self-consistent iteration scheme.

Numerical solutions in the preceding section suggest that
the population difference $r_3$ has a periodic structure
in the range, $-1 \sim 0$.
A large central part of $r_3 = 0$ corresponds to a target state of the 1/1 mixture of
the excited and ground states, and $r_3 = -1$ of the ground state at small
adjacent regions.
We may concentrate on a region of target which starts from $r_3=-1$,
rising to a plateau region of $r_3 = 0$, then returning again to  $r_3 = -1$ region.
Two regions of $r_3= -1$ may be regarded as two end regions of a condensate.

We put by hand this structure of population difference by giving
a profile of the population difference having this behavior, for instance
\begin{eqnarray}
&&
r_3(\xi) =\frac{1}{\pi} \arctan  (\frac{\xi -\Delta/2}{d} - \frac{\pi}{2}) +
\frac{1}{\pi} \arctan  (-\frac{\xi +\Delta/2}{d} - \frac{\pi}{2}) 
\,,
\end{eqnarray}
with the length of transitional  region $d $ much smaller
than the size of excited region $\Delta$; $d \ll \Delta$.
A precise form of this profile is not important in the following discussions.
As shown in Fig(\ref {single soliton}), we expect that a typical case of potential well,
$- (\gamma_+ + \gamma_- r_3)/2$ in the diagonal part of
the matrix $-{\cal M}$ of eq.(\ref{profile eq in e}),
 is close to the square well form.
For existence of bound states of a potential of this type,
it is necessary to have $\gamma_- > 0$, which
means $\alpha_{ee} > \alpha_{gg}$ ($\alpha_{ab}$ = polarizability defined in 
Appendix) implying that the excited level $|e\rangle $
must have a larger polarizability than the lower level $| g\rangle$.
For dipole transition moments of similar magnitudes, this usually holds due to
the closeness of  $|e\rangle $ to the E1 related upper level  $|p\rangle $,
closer than $|g \rangle $ to $|p\rangle $.
This profile may be extended by summing over equally displaced at $\xi = N \Delta/2 \,, N= \pm 1, \pm 2 \cdots$
in the target.
We call this the potential well model.

The classical particle motion corresponding to this potential well  is described by
\begin{eqnarray}
&&
R'' - \frac{R'^2}{2R} = - \frac{\partial}{\partial R}V
\,,
\\ &&
L'' - \frac{L'^2}{2L} = - \frac{\partial}{\partial L}V
\,,
\\ &&
V(R,L; \xi) = \frac{1}{2} (\gamma_+ + \gamma_- r_3(\xi) - 2h^2 ) (R^2+ L^2)
- 2l^2 (\ln R + \ln L)  + 4lh (R-L)
\,.
\end{eqnarray}
For $\gamma_+ + \gamma_- r_3(\xi) - 2h^2 > 0$ (for any $r_3$) 
the potential $V$ has a minimum away from the origin of $(R, L) = (0,0)$.
For $l = 0$ the potential $V$ has a form of 2d harmonic oscillator
of variable frequency given by the space dependence of $r_3(\xi)$.
This model breaks the translational invariance, hence
the zero energy solution does not exist in this case.

Semi-analytic results for the case $l=0$ of the potential model
can be worked out in the following way.
The relevant equation for this case is
\begin{eqnarray}
&&
- I'' + \frac{I'^2}{2I} + W(\xi) I = 0
\,,
\\ &&
W (\xi) = 2h^2 -  \left( \gamma_+ + \gamma_- r_3(\xi)
\right)
\,,
\end{eqnarray}
for $I = R, L$.
(In the precise non-linear equation below they are coupled by
the formula $r_3$ given by the two energy densities.)

This equation may be linearized by a change of function,
\begin{eqnarray}
&&
- \psi'' + \frac{1}{2} W(\xi) \psi = 0
\,, \hspace{0.5cm}
\psi = \sqrt{R}
\,.
\end{eqnarray}
The problem is thus equivalent to the eigenvalue problem
of Schr\"odinger equation, with the potential being
$\propto W(\xi)- h^2$ (not $V$) and the energy $ h^2 $.
It is known from textbooks of quantum mechanics that
there are always bound states  confined within a potential well
in one space dimension.
If one approximates the potential by a square well,
the number of  bound states is $n$ for 
\begin{eqnarray}
&&
   (n-1) \pi  < \sqrt{\gamma_-} \Delta  <  n \pi
\,.
\end{eqnarray}
In another word the number of bound levels is $\sim \sqrt{\gamma_-} \Delta/\pi$
for a given width $\Delta$ of the potential well.
It is important to realize that the number of bound levels is
given by a factor $\propto \Delta$ for a given target atom,
which is the size of maximally coherent excited region.
(The potential depth $\gamma_-$ proper to a chosen target plays a minor role.)
The bound levels are labeled by $k = 1, 2, \cdots n-1, n$ 
in the order of increasing energies, 
the deepest having an energy $\sim -\Delta + (\pi/\Delta)^2$ and
$k-$th excited energy  $\sim -\Delta +k^2  ( \pi/\Delta)^2$.
Excited levels are as if they had kinetic energies of momentum $\sim k$,
almost freely moving within the potential well except at the classical turning points.
There are $k-1$ nodes of the wave function for the $k-$th level, hence 
the same number of zeros for the energy density.

More generally beyond the square well model, wave functions in one-dimensional quantum mechanics
are well described by the WKB formula of quantum mechanics  for
highly excited levels.
Wave function profiles in the WKB formula
are plane-wave like near the center of the potential well,
and exponentially decreasing beyond the classical turning points
given by $W = 0$.
Almost periodic structure of R and L energy density profiles
observed numerically in previous figures of multiple soliton condensates
reflects the plane wave nature of WKB formula in the central
target region.
The presence of the exponential factor for penetration into the
potential barrier beyond the excited target region has an important consequence of suppressing
the leakage of energy flux from target ends once soliton-condensates
are created.

For a large quantum number the results of linearized potential well model support the picture in
the previous section of weakly interacting $k $ solitons
for the $k-$th bound state.
Each portion between two adjacent zeros gives a  weakly interacting soliton 
of  almost the same size $\approx \Delta/k$
(to be multiplied by $ct_0$ for the real length).
For a large potential well depth $\Delta$
there are many bound levels of different $k$, and correspondingly
solitons of different sizes.
These different soliton sizes in the real unit using the length of coherent
region $L$ may be estimated as  
\begin{eqnarray}
&&
( \Delta, \frac{\Delta}{2}, \frac{\Delta}{3
},\cdots \frac{\pi}{\sqrt{\Delta}} ) \cdot ct_0
\,, \hspace{0.5cm}
\Delta = \frac{L}{ct_0}
\,,
\end{eqnarray}
by using the square well potential.
The ground state corresponding to $k=1$ may have a gigantic linear size $\sim L$,
and may be the most effective to RENP, giving a large activity factor $\eta$.
Thus the potential well model gives a more precise picture
of soliton-condensates  complementary to previous discussions
based on numerical computations.

We did a sample numerical computation using the linearized
potential well model, and found a reasonable agreement
with numerical solution in the preceding section.
Note that the field profile of Fig(\ref{single soliton}) looks like
the wave function squared of the ground state in a
square well type of the potential  $-\gamma_- (1+r_3)/2$,
except the target region near two edges due to
that the boundary condition for the finite size
is not taken into account properly.

Stability of soliton-condensates is obvious in the
bound state picture of the potential well model.
Their bound state energies are real without 
dissipative effect.
The topological quantum number of absorber and emitter
solitons alluded above plays a minor role in
the stability analysis of solitons.
What is more important is the existence of many bound states
for a large potential well width $\Delta$.

Needless to say, there exist scattering states of almost constant envelope amplitudes
which give propagating fields modified by refractive
index of medium.
There may be resonance states having
imaginary parts, which are unstable having finite lifetimes, too.
These seem to be interpreted as corresponding to
self-induced transperancy (mentioned in \cite {coherent light propagation in 2 level}
for E1 transitions)
and explosive PSR events \cite{yst pra}.

\vspace{0.5cm}
{\bf Formulation of non-linear eigenvalue problem
for $l=0$
}

We shall formulate the non-linear eigenvalue equation for
the special case of $l=0$, relegating more complicated
cases of $l \neq 0$ to the last.

In the preceding discussion we assumed a form of population
difference $r_3(\xi)$ by heuristic arguments based on
numerical solutions in the preceding section.
We now take the correct form of the population difference
in terms of the field powers;
\begin{eqnarray}
&&
r_3 = - \frac{ 1+ 4\tau_2^2\gamma_-  ( |\psi_R|^2 +   |\psi_L|^2 )^2  }
{ 1 + 16 \tau_1 \tau_2  |\psi_R|^2|\psi_L|^2 + 4\tau_2^2 \gamma_-^2 
(  |\psi_R|^2 +  |\psi_L|^2  )^2 }
\,.
\label {population difference by field powers}
\end{eqnarray}
This quantity is in the range of $0 > r_3 \geq -1$.
The non-linear eigenvalue equation to be solved is then
a coupled set of two-component equation,
\begin{eqnarray}
&&
- \frac{d^2} {d\xi^2} \psi_a + \frac{1}{2} \left(
 h^2 - \frac{1}{2} ( \gamma_+ + \gamma_- r_3(|\psi_R|^2, |\psi_L|^2)
\right) \psi_a = 0
\,,
\label {non-linear schroedinger}
\end{eqnarray}
with $a= R, L$.
The normalization of wave functions $\psi_a$ is given
by the integrated sum of stored energies, both from
light fields and medium excitation.
The assumption of this type of
normalization condition is sound on physical grounds, but
we do not know its precise form at the moment.
The exact conservation law valid for finite relaxation times $T_i$
given in \cite{yst pra} is not useful because it is based
on the first order formalism of field envelope equation
unlike the second order formalism here.

When one of the fields is absent ($\psi_R=0$ or $\psi_L=0$), $r_3 = -1$ automatically
and the non-linear potential in eq.(\ref{non-linear schroedinger})
is a mere constant.
This case does not give any bound state solution, hence
there is no soliton condensate in the one-field case.

The linearized model of potential well in the preceding
discussion may be taken as the first trial approximation
in the self-consistent iteration scheme.
Result of  wave function obtained by this linear problem
is then inserted into the formula   eq.(\ref {population difference by field powers}) 
for the population difference.
The second round of approximation assumes this function 
for a better choice of $r_3(\xi)$ and again solve the linear problem 
with this potential to obtain the
second order form of wave functions.
One continues this process until one reaches a reasonable
level of consistent $r_3$ and fields.
One may well call this a non-linear Schr\"odinger equation.
If the first guess for $r_3(\xi)$ is good, one may expect a good convergent series of
wave functions $\psi_a$ and eigenvalues in several steps
of iteration.
We cannot rigorously prove the convergence of iteration, but
it seems that the non-linear eigenvalue problem proposed here is physically
a sound formulation.

\vspace{0.5cm}
{\bf 
General non-linear eigenvalue equation for $l\neq 0$
}

In the most general case the harmonic force $W$ has
further terms dependent on $l$;
\begin{eqnarray}
&&
- I'' + \frac{I'^2}{2I} + W_I I = 0
\,,
\label {classical analogue eq}
\\ &&
W_I =  - \frac{1}{2} \left( \gamma_+ + \gamma_- r_3 \right)
+ 2 \frac{(l \mp h I)^2}{I^2} \,.
\end{eqnarray}
$\mp$ corresponding to the choice of $I = R, L$.
The non-linear Schroedinger equation becomes
more complicated,
\begin{eqnarray}
&&
- \frac{d^2} {d\xi^2} \psi_a
- \frac{1}{4} ( \gamma_+ + \gamma_- r_3(|\psi_R|^2, |\psi_L|^2) \psi_a
+ \frac{ (l \mp h |\psi_a|^2)^2}{ |\psi_a|^4} \psi_a= 0
\,.
\label  {non-linear schroedinger 2}
\end{eqnarray}

The last term in the non-linear Schr\"odinger equation (\ref {non-linear schroedinger 2})
is bounded from below, and it may be reasonable to
take wave functions around its minimum.
A constant term around this minimum exists
in the equation.
One may regard this as freely propagating light field
that does not contribute to condensates.
Deviation from this free propagation
makes up created condensates.

Non-linear terms that appear in this general case
are somewhat awkward looking and from the elegance point
it may be better to go back to the original set of
real four component (in terms of $e_i, i= 1 \sim 4$) equation, eq.(\ref {profile eq in e}).
This is another precise form of non-linear eigenvalue equation of elegance,
treating all Bloch vector components $r_i\,, i=1,2,3$ on an equal footing:
the sacrifice for elegance is the loss of conserved quantities, $h,l$,
along with a practical usefulness (presumably).
In the four-component non-linear eigenvalue formalism
it is evident that effects of relaxation with finite $T_i$'s
are all condensed in diminished magnitudes of Bloch
vector $\vec{r}^2 < 1$ and nowhere else at all.

A remaining important problem
is to determine the probability distribution
of various forms of solitons, in particular their
size distribution,  created
as a result of time dependent evolution of
laser irradiation along with events emitted as
PSR, which is left to future work.

\vspace{0.5cm}
In summary,
we demonstrated for the first time
that despite of existence of relaxation effects
stable soliton-condensates
exist in medium atomic system coupled to
two-photon mode.
There are a variety of condensate profiles
depending on strength of soliton-soliton interaction,
which have been identified by numerically solving exact
non-linear profile equations for two reduced dynamical variables
in infinitely long targets.
Single solitons of unit topological charge
and their condensates of net zero charge
are obtained this way.
For a given finite size target a precise non-linear
eigenvalue equation has been formulated, to give the profile of
condensates as wave functions of quantized bound states.
The first iterative approximation to this non-linear system is equivalent to
a linearized eigenvalue problem in one space dimension.
We were able to identify condensates of a large number of solitons
as highly excited bound levels in a potential well of large size.
The leakage of fluxes from target ends is found to be
exponentially small, being given by the penetration factor
into potential barriers.
Dynamical process of formation
of soliton-condensates ought to
be worked out by solving the full space-time evolution
equation in order to further determine the soliton size distribution.

\vspace{0.5cm}
\lromn5 
{\bf Appendix:  Second order master equation
and basic scale units}

The basic time and length unit $t_0$ and $ct_0$
are identified from the Maxwell equation
$(\partial_t^2 - \partial_x^2) E = - \partial_t^2 P$.
RHS of the Maxwell equation has a off-diagonal element
of the polarizability matrix $\alpha$ later given, $\alpha_{ge}$
as the coupling parameter and linear both in Bloch
vector components $R_i = nr_i$ and $E$.
We thus identify the basic units and dimensionless
time $\tau$ and length $\xi$ as
\begin{eqnarray}
&&
t_0 = (\frac{1}{2} \epsilon_{eg} \sqrt{\alpha_{ge}n})^{-1}
= \sqrt{\alpha_{ge}n} t_*
\,, \hspace{0.5cm}
\tau =  \frac{t}{t_0}
\,, \hspace{0.5cm}
\xi = \frac{x}{ct_0}
\,,
\end{eqnarray}
where $t_*$ is the unit defined in PTEP paper
\cite{ptep overview}.

The field strength unit is identified from the
Bloch equation, which is first order in time derivative
and in RHS contains $\alpha_{ge} E^2 R_j$.
Hence the unit field strength $E_0^2$ and dimensionless
quantity $e_i$ are defined by
\begin{eqnarray}
&&
E_0^2 = (t_0 \alpha_{ge}/2 )^{-1}
\,. \hspace{0.5cm}
e_i = \frac{E_i}{E_0}
\,.
\end{eqnarray}
These modifications of basic units give 
scale dimensions when solitons are formed.
In particular, length and time units
are considerably smaller than those of \cite{ptep overview}.
The RENP rate unit $\propto E_0^2 L$ ($L=$ the target length) is unchanged 
and coincides with previous results,
because the scale change from \cite{ptep overview}  remains this combination invariant.
The dynamical factor, in particular, after solitons are formed,
is calculated using the soliton profile obtained
by solving the profile equation written in the new units. 

In the example of para-H$_2$ vibrational transition
$v=0 \leftrightarrow 1$,
\begin{eqnarray}
&&
\alpha =
\left(
\begin{array}{cc}
1.1 & 0.069 \\
0.069 & 1.0
\end{array}
\right)
\times 10^{-23} {\rm cm}^{3}
\,,
\end{eqnarray}
at $\omega_1 = \omega_2 = \epsilon_{eg}/2\,, 
\epsilon_{eg} \sim 0.52{\rm eV}$.
We converted these values in \cite{ptep overview} to
those in the relevant rationalized unit.
This gives the basic units of
\begin{eqnarray}
&&
c t_0 \sim 0.03 {\rm mm} (\frac{n}{10^{21}{\rm cm}^{-3}})^{-1/2}
\,, \hspace{0.5cm}
E_0^2 \sim 1 {\rm TW mm}^{-2}(\frac{n}{10^{21}{\rm cm}^{-3}})^{1/2}
\,.
\end{eqnarray}

The master equation in PTEP paper \cite{ptep overview}
has been written  
by using with SVEA the first order
Maxwell equation.
This is modified in the second order formalism, with the change 
$-\partial_t^2 \rightarrow \omega^2$ in RHS of Maxwell equation.
The master equation at the middle point of frequency
$\omega=\epsilon_{eg}/2$ and ignoring grating effects
is then given by
\begin{eqnarray}
&&
\partial_{\tau} r_T = -2 i \gamma_- (|e_R|^2 + |e_L|^2) r_T
+ 4 i (e_R e_L)^* r_3 - \frac{r_T}{\tau_2}
\,,
\\ &&
\partial_{\tau} r_3 = - 4 \Im (e_R e_L r_T)
- \frac{r_3 + 1}{\tau_1}
\,,
\\ &&
(\partial_{\tau}^2 - \partial_{\xi}^2) e_R = 
\frac{1}{2} \left( (\gamma_+ + \gamma_- r_3 )e_R
+ r_T^* e_L^*
\right)
\,,
\\ &&
(\partial_{\tau}^2 - \partial_{\xi}^2) e_L = 
\frac{1}{2} \left( (\gamma_+ + \gamma_- r_3 )e_L
+ r_T^* e_R^*
\right)
\,.
\end{eqnarray}
$\gamma_+ \sim 15, \gamma_- \sim 0.64$ for para-H$_2$.

\vspace{0.5cm}
{\bf Acknowledgements}
\hspace{0.2cm}
This research was partially supported by Grant-in-Aid for Scientific
Research on Innovative Areas "Extreme quantum world opened up by atoms"
(21104002)
from the Ministry of Education, Culture, Sports, Science, and Technology.

\end{document}